\input harvmac

\let\includefigures=\iftrue
\let\useblackboard=\iftrue
\newfam\black

\includefigures
\message{If you do not have epsf.tex (to include figures),}
\message{change the option at the top of the tex file.}
\input epsf
\def\figin{\epsfcheck\figin}\def\figins{\epsfcheck\figins}
\def\epsfcheck{\ifx\epsfbox\UnDeFiNeD
\message{(NO epsf.tex, FIGURES WILL BE IGNORED)}
\gdef\figin##1{\vskip2in}\gdef\figins##1{\hskip.5in}
\else\message{(FIGURES WILL BE INCLUDED)}%
\gdef\figin##1{##1}\gdef\figins##1{##1}\fi}
\def\DefWarn#1{}
\def\figinsert{\goodbreak\midinsert}
\def\ifig#1#2#3{\DefWarn#1\xdef#1{fig.~\the\figno}
\writedef{#1\leftbracket fig.\noexpand~\the\figno}%
\figinsert\figin{\centerline{#3}}\medskip\centerline{\vbox{
\baselineskip12pt\advance\hsize by -1truein
\noindent\footnotefont{\bf Fig.~\the\figno:} #2}}
\endinsert\global\advance\figno by1}
\else
\def\ifig#1#2#3{\xdef#1{fig.~\the\figno}
\writedef{#1\leftbracket fig.\noexpand~\the\figno}%
\global\advance\figno by1}
\fi
%

\useblackboard
\message{If you do not have msbm (blackboard bold) fonts,}
\message{change the option at the top of the tex file.}
\font\blackboard=msbm10 scaled \magstep1
\font\blackboards=msbm7
\font\blackboardss=msbm5
\textfont\black=\blackboard
\scriptfont\black=\blackboards
\scriptscriptfont\black=\blackboardss

\else

\fi
%
\def\subsubsec#1{\bigskip\noindent{\it{#1}} \bigskip}
\def\yboxit#1#2{\vbox{\hrule height #1 \hbox{\vrule width #1
\vbox{#2}\vrule width #1 }\hrule height #1 }}
\def\fillbox#1{\hbox to #1{\vbox to #1{\vfil}\hfil}}
\def\ybox{{\lower 1.3pt \yboxit{0.4pt}{\fillbox{8pt}}\hskip-0.2pt}}
%
%


\def\comments#1{}
\def\cc{{\rm c.c.}}

\def\p{\partial}

\def\half{{1\over 2}}

\def\tr{{\rm tr\ }}
\def\Re{{\rm Re\hskip0.1em}}

\def\bra#1{{\langle}#1|}
\def\ket#1{|#1\rangle}

\def\vev#1{\langle{#1}\rangle}

\def\CN{{\cal N}}
\def\CO{{\cal O}}
\def\CP{{\cal P}}

\def\CZ{{\cal Z}}

\def\ap{\alpha'}

\def\II{\relax{I\kern-.10em I}}

\def\IZ{\relax\ifmmode\mathchoice
{\hbox{\cmss Z\kern-.4em Z}}{\hbox{\cmss Z\kern-.4em Z}}
{\lower.9pt\hbox{\cmsss Z\kern-.4em Z}}
{\lower1.2pt\hbox{\cmsss Z\kern-.4em Z}}
\else{\cmss Z\kern-.4emZ}\fi}
\def\IB{\relax{\rm I\kern-.18em B}}
\def\IC{{\relax\hbox{$\inbar\kern-.3em{\rm C}$}}}
\def\ID{\relax{\rm I\kern-.18em D}}
\def\IE{\relax{\rm I\kern-.18em E}}
\def\IF{\relax{\rm I\kern-.18em F}}
\def\IG{\relax\hbox{$\inbar\kern-.3em{\rm G}$}}
\def\IGa{\relax\hbox{${\rm I}\kern-.18em\Gamma$}}
\def\IH{\relax{\rm I\kern-.18em H}}
\def\II{\relax{\rm I\kern-.18em I}}
\def\IK{\relax{\rm I\kern-.18em K}}
\def\IP{\relax{\rm I\kern-.18em P}}

%

\def\ad{{\rm ad}}
\def\inbar{\,\vrule height1.5ex width.4pt depth0pt}

\def\p{\partial}

\def\pb{{\bar \p}}

\font\cmss=cmss10 
\def\IR{\relax{\rm I\kern-.18em R}}

%


%

\def\gs{g_s}
\def\lp10{\ell_p^{10}}
\def\lp11{\ell_p^{11}}
\def\R11{R_{11}}

\def\frac#1#2{{#1 \over #2}}

\global\newcount\itemno \global\itemno=0

\def\itemaut#1{\global\advance\itemno by1\noindent\item{\the\itemno. } #1}


\def\Ione{\hbox{$1\hskip -1.2pt\vrule depth 0pt height 1.53ex width 0.7pt
                  \vrule depth 0pt height 0.3pt width 0.12em$}}

\def\next#1{\subsubsec{#1}}

\def\eg{{\it e.g.}}


\newdimen\tableauside\tableauside=1.0ex
\newdimen\tableaurule\tableaurule=0.4pt
\newdimen\tableaustep
\def\phantomhrule#1{\hbox{\vbox to0pt{\hrule height\tableaurule width#1\vss}}}
\def\phantomvrule#1{\vbox{\hbox to0pt{\vrule width\tableaurule height#1\hss}}}
\def\sqr{\vbox{%
  \phantomhrule\tableaustep
  \hbox{\phantomvrule\tableaustep\kern\tableaustep\phantomvrule\tableaustep}%
  \hbox{\vbox{\phantomhrule\tableauside}\kern-\tableaurule}}}
\def\squares#1{\hbox{\count0=#1\noindent\loop\sqr
  \advance\count0 by-1 \ifnum\count0>0\repeat}}
\def\tableau#1{\vcenter{\offinterlineskip
  \tableaustep=\tableauside\advance\tableaustep by-\tableaurule
  \kern\normallineskip\hbox
    {\kern\normallineskip\vbox
      {\gettableau#1 0 }%
     \kern\normallineskip\kern\tableaurule}%
  \kern\normallineskip\kern\tableaurule}}
\def\gettableau#1 {\ifnum#1=0\let\next=\null\else
  \squares{#1}\let\next=\gettableau\fi\next}

\tableauside=1.0ex
\tableaurule=0.4pt


 %
 %
 \def\eqnn#1{\xdef #1{(\secsym\the\meqno)}\writedef{#1\leftbracket#1}%
 \global\advance\meqno by1\wrlabeL#1}
 \def\eqna#1{\xdef #1##1{\hbox{$(\secsym\the\meqno##1)$}}
 \writedef{#1\numbersign1\leftbracket#1{\numbersign1}}%
 \global\advance\meqno by1\wrlabeL{#1$\{\}$}}
 \def\eqn#1#2{\xdef #1{(\secsym\the\meqno)}\writedef{#1\leftbracket#1}%
 \global\advance\meqno by1$$#2\eqno#1\eqlabeL#1$$}


\def\eg{{\it e.g.}}
\def\ie{{\it i.e.}}

\hyphenation{Di-men-sion-al}



\lref\BershadskyZS{
M.~Bershadsky and I.~R.~Klebanov,
``Partition functions and physical states in two-dimensional quantum gravity
Nucl.\ Phys.\ B {\bf 360}, 559 (1991).
}

\lref\witten{ E.~Witten, ``Ground ring of two-dimensional string
theory,'' Nucl.\ Phys.\ B {\bf 373}, 187 (1992)
[arXiv:hep-th/9108004];
 E.~Witten and B.~Zwiebach, ``Algebraic structures and
differential geometry in 2D string theory,'' Nucl.\ Phys.\ B
{\bf 377}, 55 (1992) [arXiv:hep-th/9201056].
}

\lref\PolchinskiMH{
J.~Polchinski,
``Remarks On The Liouville Field Theory,''
UTTG-19-90
{\it Presented at Strings '90 Conf., College Station, TX, Mar 12-17, 1990}
}
\lref\PolchinskiMB{
J.~Polchinski,
``What is string theory?,''
arXiv:hep-th/9411028.
}
\lref\Berenstein{
D.~Berenstein,
``A toy model for the AdS/CFT correspondence,''
JHEP {\bf 0407}, 018 (2004)
[arXiv:hep-th/0403110].
}
\lref\GinspargIS{
P.~H.~Ginsparg and G.~W.~Moore,
``Lectures on 2-D gravity and 2-D string theory,''
arXiv:hep-th/9304011.
}
\lref\KlebanovQA{
I.~R.~Klebanov,
``String theory in two-dimensions,''
arXiv:hep-th/9108019.
}

\lref\KlebanovVP{
I.~R.~Klebanov,
``Ward identities in two-dimensional string theory,''
Mod.\ Phys.\ Lett.\ A {\bf 7}, 723 (1992)
[arXiv:hep-th/9201005].
}

\lref\KlebanovUI{
I.~R.~Klebanov and A.~Pasquinucci,
``Correlation functions from two-dimensional string ward identities,''
Nucl.\ Phys.\ B {\bf 393}, 261 (1993)
[arXiv:hep-th/9204052].
}
\lref\PolyakovQX{
A.~M.~Polyakov,
``Selftuning fields and resonant correlations in 2-d gravity,''
Mod.\ Phys.\ Lett.\ A {\bf 6}, 635 (1991).
}
\lref\MaldacenaRE{ J.~M.~Maldacena,
Adv.\ Theor.\ Math.\ Phys.\  {\bf 2}, 231 (1998) [Int.\ J.\
Theor.\ Phys.\  {\bf 38}, 1113 (1999)] [arXiv:hep-th/9711200].
}

\lref\DiFrancescoSS{
P.~Di Francesco and D.~Kutasov,
``Correlation functions in 2-D string theory,''
Phys.\ Lett.\ B {\bf 261}, 385 (1991);
P.~Di Francesco and D.~Kutasov,
``World sheet and space-time physics in two-dimensional (Super)string theory,''
Nucl.\ Phys.\ B {\bf 375}, 119 (1992)
[arXiv:hep-th/9109005].
}
\lref\KlebanovHX{
I.~R.~Klebanov and A.~M.~Polyakov,
``Interaction of discrete states in two-dimensional string theory,''
Mod.\ Phys.\ Lett.\ A {\bf 6}, 3273 (1991)
[arXiv:hep-th/9109032].
}
\lref\BoulatovXZ{
D.~Boulatov and V.~Kazakov,
``One-dimensional string theory with vortices as the upside down matrix
oscillator,''
Int.\ J.\ Mod.\ Phys.\ A {\bf 8}, 809 (1993)
[arXiv:hep-th/0012228].
}
\lref\KazakovPM{
V.~Kazakov, I.~K.~Kostov and D.~Kutasov,
``A matrix model for the two-dimensional black hole,''
Nucl.\ Phys.\ B {\bf 622}, 141 (2002)
[arXiv:hep-th/0101011].
}
\lref\tHooftJZ{
G.~'t Hooft,
``A Planar Diagram Theory For Strong Interactions,''
Nucl.\ Phys.\ B {\bf 72}, 461 (1974).
}
\lref\BrezinSV{
E.~Brezin, C.~Itzykson, G.~Parisi and J.~B.~Zuber,
``Planar Diagrams,''
Commun.\ Math.\ Phys.\  {\bf 59}, 35 (1978).
}

\lref\CorleyZK{
S.~Corley, A.~Jevicki and S.~Ramgoolam,
``Exact correlators of giant gravitons from dual N = 4 SYM theory,''
Adv.\ Theor.\ Math.\ Phys.\  {\bf 5}, 809 (2002)
[arXiv:hep-th/0111222].
}

\lref\HashimotoZP{
A.~Hashimoto, S.~Hirano and N.~Itzhaki,
``Large branes in AdS and their field theory dual,''
JHEP {\bf 0008}, 051 (2000)
[arXiv:hep-th/0008016].
}

\lref\AlexandrovQK{
S.~Y.~Alexandrov, V.~A.~Kazakov and I.~K.~Kostov,
``2D string theory as normal matrix model,''
Nucl.\ Phys.\ B {\bf 667}, 90 (2003)
[arXiv:hep-th/0302106].
}

\lref\OkounkovSP{
A.~Okounkov, N.~Reshetikhin and C.~Vafa,
``Quantum Calabi-Yau and classical crystals,''
arXiv:hep-th/0309208.
}

\lref\KontsevichTI{
M.~Kontsevich,
``Intersection theory on the moduli space of curves and the matrix Airy
function,''
Commun.\ Math.\ Phys.\  {\bf 147}, 1 (1992).
}

\lref\GaiottoYB{
D.~Gaiotto and L.~Rastelli,
``A paradigm of open/closed duality: Liouville D-branes and the Kontsevich
model,''
arXiv:hep-th/0312196.
}

\lref\rajesh{
R.~Gopakumar,
``From free fields to AdS. I, II''
arXiv:hep-th/0308184, 0402063.
}

\lref\KlebanovKM{
I.~R.~Klebanov, J.~Maldacena and N.~Seiberg,
``D-brane decay in two-dimensional string theory,''
JHEP {\bf 0307}, 045 (2003)
[arXiv:hep-th/0305159].
}

\lref\DouglasUP{
M.~R.~Douglas, I.~R.~Klebanov, D.~Kutasov, J.~Maldacena, E.~Martinec and N.~Seiberg,
``A new hat for the c = 1 matrix model,''
arXiv:hep-th/0307195.
}

\lref\TeschnerRV{
J.~Teschner,
``Liouville theory revisited,''
Class.\ Quant.\ Grav.\  {\bf 18}, R153 (2001)
[arXiv:hep-th/0104158].
}

\lref\GuptaFU{
A.~Gupta, S.~P.~Trivedi and M.~B.~Wise,
``Random Surfaces In Conformal Gauge,''
Nucl.\ Phys.\ B {\bf 340}, 475 (1990).
}

\lref\SeibergEB{
N.~Seiberg,
``Notes On Quantum Liouville Theory And Quantum Gravity,''
Prog.\ Theor.\ Phys.\ Suppl.\  {\bf 102}, 319 (1990).
}

\lref\BershadskyZS{
M.~Bershadsky and I.~R.~Klebanov,
``Genus One Path Integral In Two-Dimensional Quantum Gravity,''
Phys.\ Rev.\ Lett.\  {\bf 65}, 3088 (1990);
``Partition functions and physical states in two-dimensional quantum gravity
Nucl.\ Phys.\ B {\bf 360}, 559 (1991).
}

\lref\girvin{
S.~M.~Girvin,
``The Quantum Hall Effect: Novel Excitations and Broken Symmetries,''
Topological Aspects of Low Dimensional Systems, ed. A. Comtet, T.   Jolicoeur, S. Ouvry, F. David (Springer-Verlag, Berlin and Les Editions de   Physique, Les Ulis, 2000), [arXiv:cond-mat/9907002].}

\lref\FateevIK{
V.~Fateev, A.~B.~Zamolodchikov and A.~B.~Zamolodchikov,
``Boundary Liouville field theory. I: Boundary state and boundary  two-point
function,''
arXiv:hep-th/0001012.
}

\lref\ZamolodchikovAH{
A.~B.~Zamolodchikov and A.~B.~Zamolodchikov,
``Liouville field theory on a pseudosphere,''
arXiv:hep-th/0101152.
}
\lref\TeschnerMD{
J.~Teschner,
``Remarks on Liouville theory with boundary,''
arXiv:hep-th/0009138.
}

\lref\PolyakovTJ{ A.~M.~Polyakov,
``String theory and quark confinement,''
Nucl.\ Phys.\ Proc.\ Suppl.\  {\bf 68}, 1 (1998)
[arXiv:hep-th/9711002].
}

\lref\GopakumarKI{
R.~Gopakumar and C.~Vafa,
``On the gauge theory/geometry correspondence,''
Adv.\ Theor.\ Math.\ Phys.\  {\bf 3}, 1415 (1999)
[arXiv:hep-th/9811131].
}

\lref\winftyrefs{
J.~Avan and A.~Jevicki,
``Classical integrability and higher symmetries of collective string field
theory,''
Phys.\ Lett.\ B {\bf 266}, 35 (1991);
D.~Minic, J.~Polchinski and Z.~Yang,
``Translation invariant backgrounds in (1+1)-dimensional string theory,''
Nucl.\ Phys.\ B {\bf 369}, 324 (1992);
G.~W.~Moore and N.~Seiberg,
``From loops to fields in 2-D quantum gravity,''
Int.\ J.\ Mod.\ Phys.\ A {\bf 7}, 2601 (1992);
S.~R.~Das, A.~Dhar, G.~Mandal and S.~R.~Wadia,
``Gauge theory formulation of the C = 1 matrix model: Symmetries and discrete
states,''
Int.\ J.\ Mod.\ Phys.\ A {\bf 7}, 5165 (1992)
[arXiv:hep-th/9110021].
}

\lref\AvanKQ{
J.~Avan and A.~Jevicki,
``Classical integrability and higher symmetries of collective string field
theory,''
Phys.\ Lett.\ B {\bf 266}, 35 (1991).
}
\lref\MinicRK{
D.~Minic, J.~Polchinski and Z.~Yang,
Nucl.\ Phys.\ B {\bf 369}, 324 (1992).
}
\lref\MooreAG{
G.~W.~Moore and N.~Seiberg,
Int.\ J.\ Mod.\ Phys.\ A {\bf 7}, 2601 (1992).
}
\lref\DasQB{
S.~R.~Das, A.~Dhar, G.~Mandal and S.~R.~Wadia,
``Gauge theory formulation of the C = 1 matrix model: Symmetries and discrete
states,''
Int.\ J.\ Mod.\ Phys.\ A {\bf 7}, 5165 (1992)
[arXiv:hep-th/9110021].
}

\lref\screening{
B.~L.~Feigin and D.~B.~Fuchs, unpublished;
V.~S.~Dotsenko and V.~A.~Fateev,
``Conformal Algebra And Multipoint Correlation Functions In  2d Statistical
Nucl.\ Phys.\ B {\bf 240}, 312 (1984);
``Four Point Correlation Functions And The Operator Algebra In The
Two-Dimensional Conformal Invariant Theories With $c<1$,''
Nucl.\ Phys.\ B {\bf 251}, 691 (1985).
}

\lref\DiFrancescoNK{ P.~Di Francesco, P.~Mathieu and D.~Senechal,
``Conformal field theory,'' New York, USA: Springer (1997) 890 p.
}

\lref\GoulianQR{
M.~Goulian and M.~Li,
``Correlation Functions In Liouville Theory,''
Phys.\ Rev.\ Lett.\  {\bf 66}, 2051 (1991).
}

\lref\GrossTU{
D.~J.~Gross,
``Two-dimensional QCD as a string theory,''
Nucl.\ Phys.\ B {\bf 400}, 161 (1993)
[arXiv:hep-th/9212149].
}

\lref\stringyexclusion{
J.~M.~Maldacena and A.~Strominger,
``AdS(3) black holes and a stringy exclusion principle,''
JHEP {\bf 9812}, 005 (1998)
[arXiv:hep-th/9804085];
S.~S.~Gubser,
``Can the effective string see higher partial waves?,''
Phys.\ Rev.\ D {\bf 56}, 4984 (1997)
[arXiv:hep-th/9704195];
J.~Polchinski,
``Classical limit of (1+1)-dimensional string theory,''
Nucl.\ Phys.\ B {\bf 362}, 125 (1991).
}

\lref\McGreevyCW{
J.~McGreevy, L.~Susskind and N.~Toumbas,
``Invasion of the giant gravitons from anti-de Sitter space,''
JHEP {\bf 0006}, 008 (2000)
[arXiv:hep-th/0003075].
}

\lref\WittenZD{
E.~Witten,
``Ground ring of two-dimensional string theory,''
Nucl.\ Phys.\ B {\bf 373}, 187 (1992)
[arXiv:hep-th/9108004].
}

\lref\SeibergNM{ N.~Seiberg and D.~Shih,
``Branes, rings and matrix models in minimal (super)string theory,''
JHEP {\bf 0402}, 021 (2004) [arXiv:hep-th/0312170].
}

\lref\McGreevyKB{
J.~McGreevy and H.~Verlinde,
``Strings from tachyons,''
JHEP {\bf 0312}, 054 (2003)
[arXiv:hep-th/0304224].
}
\lref\McGreevyEP{
J.~McGreevy, J.~Teschner and H.~Verlinde,
``Classical and quantum D-branes in 2D string theory,''
JHEP {\bf 0401}, 039 (2004)
[arXiv:hep-th/0305194].
}

\lref\DouglasXV{
M.~R.~Douglas,
``Conformal field theory techniques for large N group theory,''
arXiv:hep-th/9303159.
}

\lref\AganagicQJ{
M.~Aganagic, R.~Dijkgraaf, A.~Klemm, M.~Marino and C.~Vafa,
``Topological strings and integrable hierarchies,''
arXiv:hep-th/0312085.
}

\lref\GhoshalWM{
D.~Ghoshal and C.~Vafa,
``c = 1 string as the topological theory of the conifold,''
Nucl.\ Phys.\ B {\bf 453}, 121 (1995)
[arXiv:hep-th/9506122].
}

\lref\VafaQA{
C.~Vafa,
``Two dimensional Yang-Mills, black holes and topological strings,''
arXiv:hep-th/0406058.
}

\lref\stone{
M.~Stone,
``Schur Functions, Chiral Bosons And The Quantum Hall Effect Edge
States,''
ILL-TH-90-11
}

\Title{\vbox{\baselineskip12pt\hbox{hep-th/0408180}
\hbox{PUTP/2130}}} {\vbox{ \centerline{The Large N Harmonic
Oscillator as
%
a String Theory}}}

\bigskip
\centerline{Nissan Itzhaki and John McGreevy}
\bigskip
\centerline{{\it Department of Physics, Princeton University,
Princeton, NJ 08544}}
\bigskip
\bigskip
\noindent We propose a duality between the  large-$N$ gauged
harmonic oscillator and a novel  string theory in two dimensions.

\bigskip
\Date{August, 2004}

\newsec{Introduction}

Since  't Hooft's work  thirty years ago \tHooftJZ \ it is
generally believed that large-$N$ gauge theories admit a dual
string theory description. Given the utility of such dualities for
both sides, and the difficulty of finding them, it is of interest
to find more examples which are under precise control.
The aim of this paper is to explore a
new and remarkably simple
duality of this kind. We address an old question that was
resurrected recently  \Berenstein : what is the string theory dual
of the large-$N$ gauged harmonic oscillator?

A clue comes from the fact that the large $N$ {\it inverted}
harmonic oscillator is dual to a certain string theory in two
dimensions (for reviews see
\refs{\KlebanovQA,\GinspargIS,\PolchinskiMB}). Therefore, we
should find the meaning in string theory of rectifying the
inverted matrix potential.
This is done in section 4, after discussing some of the properties
and symmetries of the large-$N$ harmonic oscillator in sections 2
and 3. The physics of the harmonic oscillator and the inverted
oscillator are very different. The latter has a continuous
spectrum, while in the former the spectrum is discrete. This
implies that, unlike in the standard 2d-string/matrix model
duality, there is no need to take a double scaling limit to find a
continuum dual string description. Finite $N$ corresponds to
nonzero string coupling constant. In this sense, the duality we
propose works more like the AdS/CFT duality \MaldacenaRE\ than the
standard 2d-string/matrix model duality.  Since the matrix model
is well-defined at finite $N$, we will be able to define and study
interesting finite-coupling effects in this bosonic string theory.

The observables of interest in the large-$N$ harmonic oscillator
are the overlap amplitudes between resonances. In section 5 we
explain how the resonances come  about on the string theory side.
In section 6 we calculate  these overlap amplitudes on both sides
of the duality and show that the large-$N$ limit of the harmonic
oscillator exactly agrees with the relevant sphere amplitudes on
the string theory side. In this section, we also make contact with
a normal matrix model.
In section 7 we show that the
duality passes non-trivial tests involving the $1/N$ corrections
to the leading large-$N$ behavior. Section 8 is devoted to a
heuristic picture of the duality.

In addition to providing a new example of open/closed string
duality, the relation we propose is interesting for a completely
different reason. The matrix harmonic oscillator is closely
related to the quantum hall effect (QHE) and therefore the  dual
stringy description might be useful for understanding various open
questions in the QHE, like what is the effective description of
the quantum phase transition from one plateau to the next. In
section 9 we briefly speculate on this and other possible
applications and generalizations.

\newsec{The Matrix Model}
The model we wish to study is the gauged quantum mechanics of an
$N\times N$
matrix harmonic oscillator,
\eqn\act{ S = \half \int dt ~ \tr \left( (D_0 X)^2 - X^2 \right)
.}
 The derivative $D_0 = \del_0 + [A_0, \cdot ]$ is covariant with
respect to gauged $U(N)$ conjugations $ X \mapsto \Omega X
\Omega^\dagger $. The gauge field acts as a Lagrange multiplier
that projects onto singlet states,
which in turn \BrezinSV\ describe $N$
free fermions in the harmonic oscillator potential.

This model is, of course, solvable. The possible energy levels of
a single fermion are \eqn\1{ E_j=j+\half.} Since there are $N$
fermions the vacuum energy is \eqn\2{E_{0}=\half+{3\over
2}+...+{(2N-1)\over 2}={N^2 \over 2}.} The Hilbert space is
spanned by states labelled by $N$ integers $k_n$ such that
 $0\leq k_1<k_2<...<k_N$, and the eigenvalues of the Hamiltonian are
\eqn\energy{H\ket{k_1,k_2,...,k_N}=
\left( {N\over 2} +\sum_{n=1}^{N}k_n\right) \ket{k_1,k_2,...,k_N}.}

Despite the fact that this model is free something quite interesting
is happening in the large-$N$ limit:
excitations above the ground state
are most easily described in terms of a chiral boson,
as emphasized recently in \Berenstein.
A nice way to see this is to consider the partition function
\eqn\partition{Z=\tr q^H=q^{N\over 2}\sum_{k_1=0}^{\infty}
\sum_{k_2=k_1+1}^{\infty}...\sum_{k_N=k_{N-1}+1}^{\infty}q^{\sum_{n=1}^{N}k_n}.
  }
Performing the sums sequentially we get \BoulatovXZ\
\eqn\partitionb{Z=q^{N^2/2}\prod_{n=1}^{N}{1\over 1-q^n}.} This is
exactly the partition function of a two-dimensional chiral boson
with $\alpha_0=N$ whose excitations are truncated at level $N$.
Namely the Hamiltonian is \eqn\hamiltonianb{ H={\alpha_0^2 \over
2}+ \sum_{n=1}^{N} \alpha_{-n}\alpha_{n},} and we are using stringy
conventions for the commutators \eqn\com{ [\alpha_m,\alpha_n]=n
\delta_{n+m}.} Later on we shall argue that this chiral boson is
(up to normalization) equivalent to the target-space field of the
string description.
Note that for finite $N$, the momentum modes of the chiral boson
are truncated in a clean way. This is another example of the
stringy exclusion principle \refs{\stringyexclusion, \GrossTU},
about which we will say more in section 5.


To understand this boson description in more detail, we introduce
matrix raising and lowering operators
\eqn\creation{ a^i_j = {1
\over \sqrt 2} \left(X^i_j + i P^i_j\right), ~~~~~ a^{\dagger~i}_j
= {1 \over \sqrt 2} \left(X^i_j - i P^i_j\right), }
 where $P$ is
the momentum conjugate to the matrix $X$. These operators satisfy
\eqn\alllgg{ [a^i_j, a^{\dagger~k}_l ] = \delta^i_l \delta^k_j,
~~~~~ [H, a] = - a, ~~~~~ [H, a^\dagger] = a^\dagger .}
The vacuum is defined by $ a^i_j \ket{0} = 0 $.
The states
\eqn\ketmi{\ket{ \{ m_i\} } \equiv c_{\{m_i\}}
\prod_{i=1}^r \tr ( a^{\dagger~m_i } ) \ket{0} ,}
with  $m_1 \geq m_2 \geq ... \geq m_r$ provide a useful
basis
for the Hilbert space. From \alllgg \ we see that
indeed the spectrum is evenly-spaced. Note that the stringy
exclusion is the statement that \eqn\excl{m_i \leq N}
 in order that these states be linearly independent.

An orthonormal basis of states can be constructed from \ketmi\ as
%
\eqn\qoq{ \ket{ R(\{ m_i\}) } = c_{\{m_i\}}~\chi_{R(\{m_i\})}
(a^\dagger) \ket{0} }
(see \eg\ \CorleyZK\ or \stone \foot{We thank D. Berenstein for
pointing out a misstatement in an earlier version, and the latter reference.})
where $R( \{ m_i \} )$ is the representation
of $U(N)$ corresponding to the Young tableau with columns of
lengths $ ( m_1, m_2, ..., m_r) $ and $\chi_R(U)$ is the character
of $U$ in this representation. The bound $m_i \leq N$ guarantees
that this is indeed a tableau for a $U(N)$ representation.

In terms of the free-fermion description, the wavefunction for the
state \qoq\ is a Slater determinant constructed as follows. Look
at the tableau $\{m_i\}$ sideways, define $k_n$ to be the
row-lengths; note that there are at most $N$ rows by \excl.  The
many-fermion wavefunction is then
\eqn\taio{\bra{z_1}\otimes\bra{z_2}\cdots\vev{z_N |R(\{m_i\}) } =
\det_{n,l = 1,..,N} \psi_{k_n + N - n + 1}(z_l)}
where $\psi_n(z)
= \vev{ n | z} = H_n(z) e^{ - |z|^2/2 } $ are the single-particle
harmonic-oscillator wavefunctions. For example, for the empty
tableau, this fills the lowest $N$ energy levels with fermions.
The state $ \tr ( a^{\dagger})^k \ket{0}$ corresponds to exciting
$k$ fermions by one level each -- it makes a hole in the fermi sea
at level $N-k$. For $m = N$, the hole is at the lowest state. For
$m> N$, there is no such single-particle description, in accord
with the stringy exclusion principle.


This is the same Hilbert space as that of two-dimensional
Yang-Mills theory on a cylinder (see \eg\ \DouglasXV \ ). The
Hamiltonian for the matrix oscillator is, however, the number of
boxes in the tableau,
\eqn\pqc{ (H - E_0)\ket{ \{ m_i\} } = \sum_{i=1}^r  m_i \ket{ \{
m_i\} }}
rather than the second Casimir $C_2(R)$.

\newsec{Symmetries}

As is well-known (see \eg\ \S12 of \GinspargIS) there are
infinitely many conserved charges associated with the harmonic
oscillator. These charges generate the $w_{\infty}$ algebra that
controls much of the physics of the system, and was studied quite
intensively in the case of the inverted harmonic oscillator
\winftyrefs\ in relation with the $c=1$ theory.
Although most of our discussion can be obtained from these papers
by plugging $i$'s in the right places, we find it useful to be
explicit
because the harmonic oscillator
physics is quite different from the inverted
oscillator physics.

At large $N$, a semi-classical description
exhibits most of the relevant
physics in a simple way. Semi-classically, the eigenvalues form a
Fermi sea in phase space.
 It is most convenient to parametrize the phase space by
\eqn\nj{U={(X+iP)\over\sqrt2},~~\;\;\;\;V=U^{\dagger}=
{(X-iP)\over \sqrt2},} which satisfy the Poisson bracket \eqn\pb{
\{ U,V\} _{PB}=i.} The Hamiltonian is $H=UV$, and thus
\eqn\eqm{U(t)=e^{-i t} U(0),\;\;\;V(t)=e^{it} V(0),} which implies
that following charges are conserved \eqn\wi{Q_{n,m}=e^{i(n-m)t}
U^n V^m.} These charges form the $w_{\infty}$ algebra \eqn\we{
\{Q_{n,m},Q_{n^{'},m^{'}}\} _{PB}
 =i(nm^{'}-mn^{'})~ Q_{n+n^{'}-1,m+m^{'}-1}.}
The ground state is obtained by filling
states in the phase space up to the Fermi surface
\eqn\gs{ UV = N,}
so that  the area in units of the Poisson brackets \pb\ is equal
to the number of fermions.

Excitations above the ground state can be described using
area-preserving transformations of the phase space. The area is
preserved since it corresponds to the number of fermions.
Area-preserving transformations can be described using a single scalar
function $h(U,V)$,
a basis for which are
\eqn\area{ h_{nm}=U^n V^m,} $n$ and $m$ must be integers in order
to preserve the connectivity of the fermi surface. $h_{nm}$
determines a vector field that is associated with
 an infinitesimal area-preserving transformation of the
U-V plane, \eqn\vecr{  \vec{B} _{nm}={ \partial h_{nm}\over
\partial U}
\partial_V- {\partial h_{nm}\over\partial V} \partial_U.}
The Lie brackets of the $\vec{B}_{nm}$'s form a $w_{\infty}$
algebra.

The
vector field $\vec{B}_{nm}$ clearly depends both on $n$ and $m$.
However, when acting on the ground state \gs, $\vec{B}_{nm}$
depends only on $|s|=|n-m|$. Let us see how this comes about. Eq.\
\vecr\ implies that $h_{nm}$ generates the following infinitesimal
deformation
\eqn\trr{\delta V=\epsilon\partial_U h=
\epsilon nU^{n-1}V^m,\;\;\;\delta U=-\epsilon
\partial_V h =-\epsilon mU^n V^{m-1}. }
Therefore, to leading order in $\epsilon$, we find that
\eqn\kl{(V+\delta V)(U+\delta U)= UV +\epsilon(n-m)U^n V^m. } If
$UV$ is to begin with a constant (like in the ground state) then
\kl $\; $implies that the deformation of the Fermi surface is
\eqn\df { \delta( UV ) \sim \epsilon  \left( {U \over V}\right)
^{s/2}.} The only dependence on $r=n+m$ is in the numerical
constant. Parameterizing  the phase space  using polar coordinates
(see fig. 1)
\eqn\polar{ U=re^{i\theta},\;\;\;V=re^{-i\theta},}
we can  write the variation in the fermi level \df\ as
\eqn\kll{ \delta(\theta)\sim \epsilon \Re (e^{is \theta })
=\epsilon \cos (|s | \theta ).}
%
If we act more than once with $\vec{B}_{nm}$ on the ground state
then both $s$ and  $r$  matter as is clear from the $w_{\infty}$
algebra. For example, both $1$ and $UV$ act trivially on the
ground state but only $1$ acts trivially on every state.
\ifig\fermisea{ For the right-side-up oscillator, the Fermi sea is
a compact droplet.  Ripples on the Fermi surface (and holes) travel
in circles with unit angular velocity. }
{\epsfxsize1.5in\epsfbox{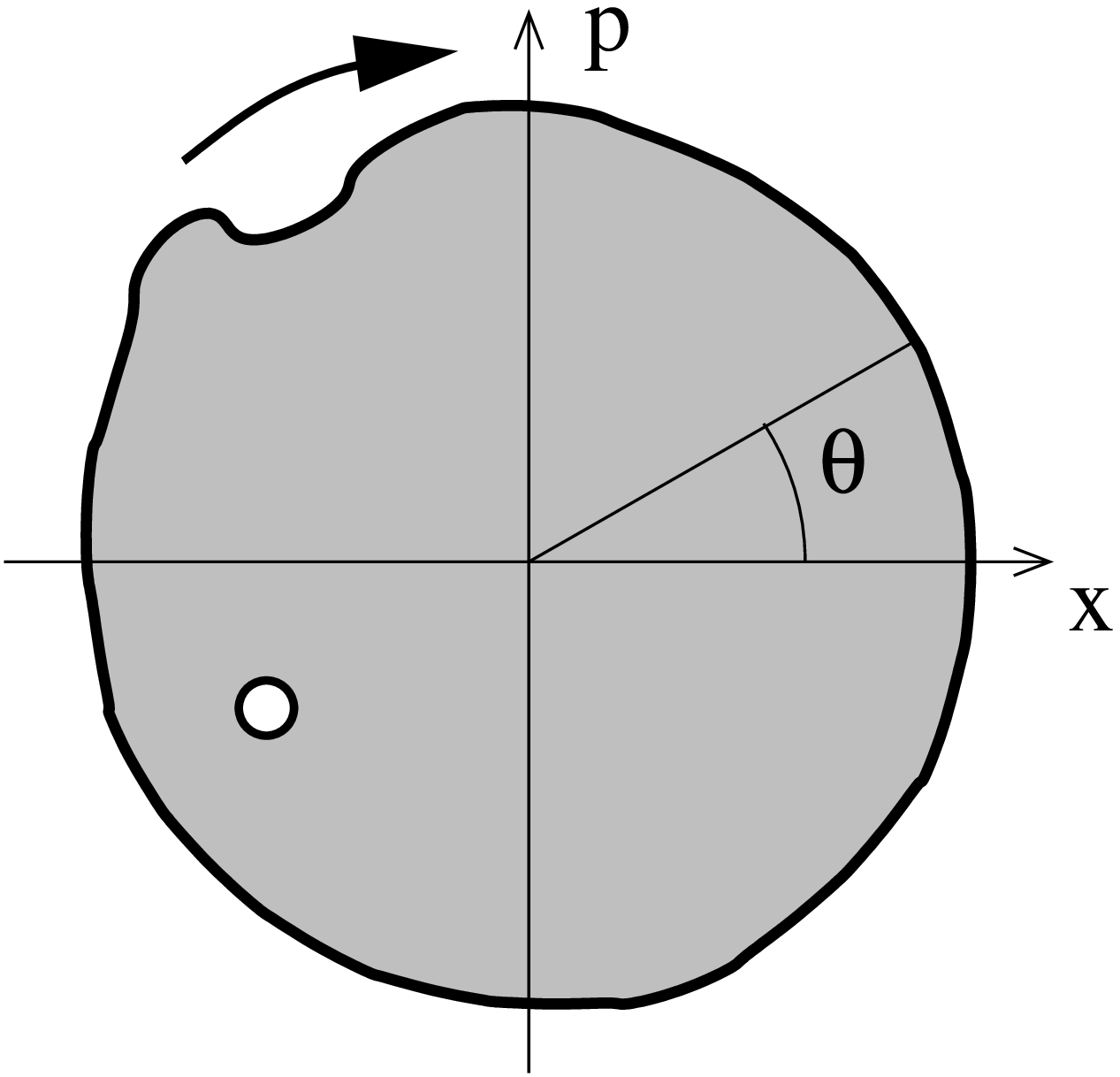}}

This can be conveniently phrased at the quantum level. Using the
bosonic description one can excite the ground state by acting with
single trace operators. For example, up to a normalization
constant, the $\alpha_{n}$ of the chiral boson, which appeared in
\hamiltonianb, correspond to \Berenstein\
   \eqn\7{\tr ((a^{\dagger})^n) ,\;\;\;\,}
with $a$ as in \creation. Momentarily we shall claim that  these
are  dual to excitations of the  closed string "tachyon". As
should be clear from the semi-classical discussion above, there
are other single trace operators in the theory that involve both
$a^{\dagger}$ and $a$. These are
associated with the  $h_{nm}$ with $m\neq 0$. For example,
consider the operator \eqn\ex{\tr (a (a^{\dagger})^2).} Acting
with this operator on the ground state has the same effect as
acting with $\tr(a^{\dagger})$. Namely, both take the ground state
to the first excited state. However clearly these operators are
not the same. In fact we will see that it is natural to relate
these
 operators to discrete states in the dual stringy description.

At this point we encounter a small puzzle. At the quantum level
there seem to be many more excitations than at the semi-classical
level, which clearly makes no sense. To see this we note that
there are many different single trace operators that at the
semi-classical level are associated with the same $h_{nm}$. The
simplest example is $ \tr (aa^\dagger a a^\dagger ) $ and $\tr
(a^2a^{\dagger~2} )$. On the one hand, since in a non-Abelian
theory the order inside the trace matters, these are  different
operators. On the other hand semi-classically they are both
associated with $h_{2,2}$. The fact that the theory is gauged
resolves this issue: in one dimension there is no electric or
magnetic field and so the current that couples to $A$ must vanish
on-shell, \foot{ Note that here we are using  matrix commutator
$[a,a^\dagger]_i^j$ which is not to be confused with  the
canonical commutator $[ a_i^j, a_k^{\dagger~l}]$ in \alllgg.}
\eqn\gauss{j=[X,D_0 X]=i[a, a^{\dagger}]=0.}
%
%

This implies that operators which differ by such commutators
actually give the same result when acting on  physical states. As
a result, the inequivalent single-trace operators are in 1-to-1
correspondence with $w_\infty$ generators. In section 5 we discuss
further these operators at the quantum level.

Eq. \gauss\ is reminiscent of a normal matrix model (for a
recent discussion in relation with the $c=1$ theory see
\AlexandrovQK ).
We will elaborate on this connection in \S6.

\newsec{The dual string theory}

Generally speaking it is not an easy task to find the stringy dual
description of some  large-$N$ gauge theory. In our case a natural
candidate  comes from the well-studied duality between 2d string
theory with a Liouville direction and matrix quantum mechanics.
The matrix quantum mechanics dual to 2d
string theory is closely related to the one that we are
considering. The kinetic term is the same and the sign
of the potential term
is flipped. Namely, the free fermions experience the celebrated
inverted quadratic potential, rather then the harmonic
oscillator potential of our case.
Starting from that
well-tested duality, what we have to do is figure out what it
means, on the string theory side, to flip the potential, and see
if what we get makes sense. In this section we discuss the meaning
of flipping  the potential on the string theory side. In the rest
of the paper we test our conjecture for the dual stringy
description.

As was emphasized in \refs{\KlebanovQA} the curvature of the
potential in the matrix model is related to the tension of the
dual string theory by
\eqn\po{ U(x)= {1 \over 2 \ap} x^2.}
So flipping the potential means that we have to take
\eqn\9{ \ap \rightarrow -\ap.}
In dimension larger than two this would cause an instability due
to the massive modes that are now tachyonic. However, in 2d there
are no massive modes (other than some discrete states that will
play an important role shortly) so we do not have that problem. A
better way to say this is that instead of \9\ what we could do is
to keep $\ap$ positive and Wick rotate {\it all} dimensions. In
$D$ dimensions  this means that there are $D-1$ time-like
directions, which causes problems if $D>2$. But in two dimensions
this just means that the previously spatial dimension is now the
time direction and vice-versa.

Let us see what happens if we apply that logic to the $c=1$
theory. Before we flip the potential the dual string theory could
be viewed as a tensor product of two CFT's, a free time-like boson
$X$ (with $c=1$), which is identified with the quantum mechanics
time, and a space-like Liouville field, $\varphi$, with $c=25$.
Namely,
%
\eqn\tt{ T(z)={1 \over \ap}\left( :\del X \del X: - :\del \varphi
 \del \varphi: \right) + {Q \over \sqrt{\ap }}\del^2
\varphi,\;\;\;\;Q=b+1/b=2. }
Adding the Liouville term, we end up with the
following worldsheet action
\eqn\wsaction{ S = {1 \over 4 \pi \ap} \int d^2 \sigma \sqrt{g} ~
\left( -\del_{\alpha} X  \del^{\alpha} X+ \del_{\alpha} \varphi  \del^{\alpha}
\varphi + \sqrt{\ap} R^{(2)} Q \varphi + \mu_0 e^{  2 b
\varphi/\sqrt{\ap}} \right).}

According to the reasoning above, to find the candidate
dual to the large-$N$ harmonic oscillator we could either apply \9\
or double Wick rotate
\eqn\wr{  X\rightarrow i X,\;\;\;\; \varphi\rightarrow i\varphi .}
Either way  we get a CFT whose stress tensor is
%
\eqn\ttt{ T(z)={1 \over \ap}\left( :\del \varphi \del \varphi:
 - :\del X  \del
X: \right) + i{Q \over \sqrt{\ap }}\del^2 \varphi .}
The energy-momentum tensor is now complex which implies that the
world-sheet theory is non-unitary. This, however, does not mean
that the target space theory is inconsistent, because of the large
redundancy of the worldsheet description. It is worth pointing out
in this context  that a similar CFT (with spatial $\varphi$ and no
$X$) is used to realize the minimal models in the Coulomb gas
formalism (for a review see \DiFrancescoNK, chapter 9). The total
central charge of \ttt\ is
\eqn\cc{c=2- 6 i^2  Q^2=26,}
where the minus sign comes from the fact that now the Liouville
direction is time-like. The worldsheet action is
\eqn\wsa{ S = {1 \over 4 \pi \ap} \int d^2 \sigma\sqrt{g}  ~
\left( \del_{\alpha} X \del^{\alpha} X - \del_{\alpha} \varphi  \del^{\alpha}
\varphi + i\sqrt{\ap} R^{(2)} Q \varphi + \mu_0 e^{ i 2 b \varphi
/\sqrt{\ap}} \right).}
Notice that in the presence of the Liouville interaction  this is
not an {\it analytic} continuation \PolchinskiMH. During the
continuation, the contour of integration of the $\varphi$ field
passes over infinitely many troughs of the Liouville potential,
and the two theories are therefore not equivalent. As is clear
from \po\ and \act, to compare this string theory with the
harmonic oscillator it is most convenient to work with $\ap =1$,
which  we use in  the rest of the paper.

The coupling of $\varphi $ to the world-sheet curvature implies
that the string coupling constant is
\eqn\scc{ g_s=e^{i 2\varphi}.}
Notice that unlike in the usual linear dilaton case there is no
separation into weakly coupled and strongly coupled regions. The
absolute value of $g_s$ is one everywhere. This seems to suggest
that there is no good expansion parameter in this theory. That is
indeed the case in the free theory, $\mu_0=0$. However, as  we
shall see in the next section, in the interacting theory $1/\mu_0$
is the parameter that controls the genus expansion.

From \scc $ $ we also see that now, again unlike the usual case,
we can compactify $\varphi $. The allowed radii of compactification
seem to be
\eqn\c{ R=m/2,}
where $m$ is an integer. However, since non-perturbatively
there are D-branes and open strings effects in the theory, the
open-string coupling constant should be well-defined. Since
$g_o^2=g_s$ we find that $m$ is an even number. So the smallest
possible radius is $1$
\eqn\c{\varphi \sim \varphi +2\pi.}
The dual string theory we propose to the large-$N$ harmonic
oscillator is \wsa $ $ with this periodicity condition \c.

A useful way to think about the radius of the $\varphi$ direction
is that  $\varphi $ is dual to $\theta$ of \polar $ $, which has
the same periodicity. The relation we are proposing between the
large-$N$ harmonic oscillator and that string theory is simple:
$X$ should be identified with the quantum mechanics time and
$\varphi$ should be identified with the angular variable,
$\theta$,  in phase-space. This seems to make sense since
excitations in both descriptions travel at the same speed ($1$ in
our units) regardless of their energy. Note, however, that on the
string theory side, at least naively, there are both left-movers
and right-movers (in the target space), while on the quantum
mechanics side there are only left-movers. In the next section we
will see how to resolve this apparent contradiction.

\newsec{Correlation functions}

In this section we describe the map between the quantum mechanics
and the string theory degrees of freedom and compare their
correlation functions.
This section is devoted to a more qualitative illustration of
the dictionary between the
two sides of the duality. More details can be found in the
next section.

Let us start with the quantum mechanics side.
The observables of interest are the overlaps between different
states
\eqn\ov{ {\cal O}({\rm bra; ket})=\vev{{\rm bra | ket }}  .}
Since the time evolution in the theory is trivial, in principle,
all other gauge invariant observables are determined by these
overlap amplitudes. The simplest case corresponds to what we
loosely speaking call
a two-point function, namely \ov\ with
\eqn\twopf{ \bra{{\rm bra }} =  \bra{0}
\tr(a^{k_1}), \;\;\;\;\;\; \ket{{\rm ket}}=\tr
((a^{\dagger})^{k_2})\ket{0}.}
By counting index lines, we see
that the leading contribution in the large-$N$
limit goes like
\eqn\rtwop{ {\cal O}(k_1;k_2) \sim \delta_{k_1,k_2} N^{k_1}.}
As usual there are $1/N$ corrections. What is special about this
large-$N$ theory is that here the $1/N $ corrections are
truncated after a finite number of terms. Namely,
\eqn\hjkp{ {\cal O}(k_1;k_2)=\delta_{k_1,k_2}\sum_{l=0}^{[k_1/2]}
C_l(k_1) N^{k_1 -2 l}, }
where the $C_l$'s depend on $k_1$ but not on $N$. Later on we
shall see how this comes about on the string theory side.

The analog of a three-point function is \ov\ with
\eqn\twopf{ \bra{{\rm bra }} =  \bra{0} \tr(a^{k}), \;\;\;\;\;\;
\ket{{\rm ket}}=\tr
((a^{\dagger})^{k_1})\tr ((a^{\dagger})^{k_2})\ket{0}.}
In that case the leading contribution in the large-$N$ limit is
\eqn\jk{ {\cal O}(k;k_1,k_2) \sim \delta_{k, k_1+k_2} N^{k-1}.}

We now turn to the string theory side.
The ghost-number-two cohomology contains
the tachyon vertex operator
\eqn\tv{T^{\pm}_k= c\bar c ~ e^{-ik (X \pm \varphi)} e^{ i 2 b
\varphi },}
where the factor $e^{ i 2 b \varphi }$ is the string coupling that
multiplies the tachyon wave function, and $k$ is an integer due to
the periodicity \c $ $. There are four kinds of modes, with the
following interpretations
$$ T^{+}_{k>0}: ~~~~{\rm incoming~~ leftmover},~~~~~~~
T^{-}_{k>0}: ~~~~{\rm incoming~~ rightmover}, $$
$$ T^{+}_{k<0}: ~~~~{\rm outgoing ~~ leftmover },~~~~~~~
T^{-}_{k<0}: ~~~~{\rm outgoing~~ rightmover}.
 $$
Note that to make the relation with the matrix model  simple, the
energy in the $X$ direction (rather than  in the $\varphi$
direction which is actually the time direction in this background)
determines in our terminology if a wave is incoming or outgoing.

S-matrix amplitudes are constructed in the usual fashion,
and take the form
\eqn\presmat{A(k_1,k_2,...,k_{n^{+}};k_{n^{+}+1},...,k_{n^{+}+n^{-}})=\int
{\cal D} X{\cal D}\varphi e^{-S} \prod_{i=1}^{n^+ + ~ n^-} \int
d^2 \sigma \sqrt{g} e^{ik_i\varphi \pm ik_i X} e^{ i 2 b \varphi}, }
where it is understood that the first $n^{+}$ tachyons have
positive chirality, and the remaining $n^{-}$ have negative chirality.
The action $S$
is given by \wsa\ (with $\ap =1$).

At first sight it seems   hard to do calculations in this theory,
since the worldsheet action contains an interaction term.
However,
we can \refs{\screening, \GoulianQR}
expand the
exponent in powers of $\mu_0$ and get
\eqn\smat{A(k_1,k_2,...,k_{n^{+}};k_{n^{+}+1},...,k_{n^{+}+n^{-}})=}
$$ \int {\cal D} X{\cal D}\varphi \sum_{n=0}^{\infty}{\mu_0^n \over
n!} (\int d^2 \sigma \sqrt{g}  e^{ i 2 b \varphi})^n e^{-S_0}
\prod_{i=1}^{n^++n^-} \int d^2 \sigma \sqrt{g} e^{ik_i\varphi \pm ik_i X} e^{
i 2 b\varphi},$$
where
\eqn\freeaction{S_0= {1 \over 4 \pi } \int d^2 \sigma\sqrt{g}  ~
\left( \del_{\alpha} X \del^{\alpha} X - \del_{\alpha} \varphi  \del^{\alpha}
\varphi + i R^{(2)} Q \varphi \right).}
Eq.\ \smat\ now takes the form of a sum over amplitudes in the
free theory with the same tachyon (or any other closed string
mode) insertions plus additional insertions of the screening
operator, $\int d^2 \sigma \sqrt{g}  e^{ i 2 b \varphi}$.

To proceed it is convenient to follow \refs{\GuptaFU, \SeibergEB,
\BershadskyZS, \GoulianQR, \PolyakovQX} and decompose $\varphi=\varphi^0
+\tilde \varphi$ and $X=X^0+\tilde X$ and integrate first the zero
modes, $\varphi^0$ and $X^0$. The integral over $X^0$ is
a delta function which
imposes energy conservation
\eqn\ec{ k_{tot}^{+} +k_{tot}^{-} = 0,~~~~~~k_{tot}^{+}
=\sum_{i=1}^{n^+ }k_i,~~~~~~k_{tot}^{-} =\sum_{i=n^{+} + 1}^{n^{+}
+ n^{-} }k_i.}
The integral over $\varphi^0$
imposes the condition
\eqn\mc{2-2g-(n+n_{+}+n_{-})+{1\over 2}(k_{tot}^{+} -k_{tot}^{-})
=0,}
where $g$ is the genus and $n$ is the number of insertions of the
screening operator.  This relation is the equivalent of momentum
conservation in the $\varphi $ direction. The last two terms need
no explanation. The rest of the  terms are usually called the
background charge, as they are induced by the coupling of
$\varphi_0$ to the integral of the world sheet curvature which is
$$ \chi=2-2g-h,$$
where $h$ is the number of holes; in our case $h$  corresponds to
the total number of closed string insertions. Combining  \mc\ and
\ec\ we get
\eqn\emc{ k_{tot}^- =-k_{tot}^+ =2-2g-(n+n_{+}+n_{-}).}
This relation implies  that $1/\mu_0 $ is indeed the genus
expansion parameter: Given a certain amplitude determined by
$k_i$, $n^+$ and $n^-$,  we see that as we increase $g$, we
decrease $n$, the power of $\mu_0$.
The phases arising from the complex dilaton conspire to make
$\mu_0^{-1}$ the string coupling constant. In fact, comparing \emc\ to
the 't Hooft counting of powers of $N$, we see that we can
identify
\eqn\ml{\mu_0 \sim N.}
Now we are ready to compare some stringy amplitudes with the
quantum results mentioned at the beginning of this section. Let us
start with $1 \rightarrow 1$ amplitudes. There are two possible
ways to satisfy the energy conservation \ec\ and momentum
conservation \mc\ conditions on the sphere ($g=0$). The first is
to take $n=0$ and $k_{tot}^- =k_{tot}^+ =0$. This can be achieved
by having two particles with the same chirality and opposite
energy (say $n^+ =2$ and $n^- =0$). Such amplitudes vanish since
they involve only  two closed string insertions, which
do not suffice
to saturate the ghost zero modes on the sphere. Indeed
there is no dual quantum mechanical amplitude for these scattering
amplitudes.

The second way to saturate \ec, \mc\ is more interesting. We take
one particle with positive chirality and energy $k$ and another
particle with the negative chirality and energy $-k$. So $n^+ =1$
and $n^- =1$. From \emc $ $ we see that $n= k$ and so the
amplitude scales like
\eqn\amp{ A(k ; -k) \sim \mu_0^k \sim N^k.}
This amplitude does not vanish since (due to the insertions of the
screening operators) it involves more than two closed string
insertions on the sphere. It is natural to make the following
identification between the closed string modes and the quantum
mechanics operators
\eqn\id{ T^+_{k>0} \Leftrightarrow \tr ((a^{\dagger})^{k})
,~~~~~~~~T^{-}_{k<0} \Leftrightarrow \tr (a^{k}).}
Indeed we see that \amp $ $ is  in agreement with the harmonic
oscillator result \rtwop $ $. Notice that amplitudes that involve
$T^+_{k<0}$ and $T^-_{k>0}$ vanish.\foot{In the next section we
shall see how this comes about in higher point functions as well.}
This resolves the puzzle arising from the naive expectation that
we should have twice as many stringy modes as excitations of the
Fermi surface, raised at the end of the previous section. Namely,
just like in the harmonic oscillator,
we only see
tachyon wavefunctions with $p_{\varphi}=-i\del_{\varphi}$ negative. In
this 'imaginary' version of Liouville theory, there is a sense in
which the Seiberg bound becomes the fact that the target-space
field is chiral.

\subsubsec{Stringy exclusion}

As usual in string theory there are corrections from
higher-genus worldsheets.
Since $n\geq 0$ we find from \emc\ non-vanishing amplitudes only for
\eqn\rl{g \leq \left[ {k \over 2} \right] ,}
which exactly agrees with \hjkp $ $.
So our string theory  has the remarkable property that the
perturbative corrections to a given amplitude truncate after a
finite number of terms. As it is clear from \rl \
there are more and more corrections as we increase  $k$.
Although the single-trace states $\tr a^{\dagger k} \ket{0}$
remain orthogonal for any $k \leq N$ (by energy conservation),
this suggests that, as in ten dimensions \McGreevyCW, the best
description of these states at $k \sim N$ may not be in terms of a
perturbative closed string. It seems likely that a geometric
D-brane mechanism may again explain the interesting finite $N$
truncation of the spectrum. In fact,   models that are closely
related to the harmonic oscillator have been considered in
\refs{\HashimotoZP, \CorleyZK} in relation with giant gravitons in
$AdS$ spaces.
A similar phenomenon of a UV cutoff on the target space momentum
of order $1/g_s$
has recently been observed in the
context of topological strings \OkounkovSP.

\subsubsec{Three-point functions}

Next we turn to ``three-point functions,'' namely, the $1\rightarrow
2$ scattering amplitudes.
From \emc\ we see that the
sphere contribution to such amplitudes scales like
\eqn\ua{A(k;-k_1,-k_2)\sim \delta_{k,k_1+k_2} \mu_0^{k-1}.}
Again we find agreement with the harmonic oscillator scaling \jk.
At the level of the discussion of this section it seems possible
to have $k_1>0$ and $k_2<0$ such that \ua $ $ appear not to
vanish. This would contradict the proposed duality. A closer look
in the next section will show that such amplitudes in fact vanish,
in agreement with the discussion below \id.

\subsec{Discrete states}

The closed-string ghost-number-two cohomology contains other
states, known as discrete states, that are the remnants of the
massive modes of the string. In the usual $c=1$ theory they appear as
non-normalizable modes with
 imaginary energy (or real Euclidean energy)
 and imaginary Liouville momentum. In our case
they become normalizable propagating modes that are as important
as the tachyon modes discussed above.

Let us review how the discrete states come about. The simplest way
to think about them is in terms of the chiral
$SU(2)$ current algebra
\eqn\su{ J^{\pm}(z)=e^{\pm 2 i X(z)}, ~~~~~J^3(z)= i \partial X(z).}
The highest weight fields with respect to this algebra are
$\Psi_{j,j}(z)=e^{2ij X}$ where $j=0,1/2,1...$. With the help of
$J^-_0=\oint {dz \over 2\pi i} e^{-2i X}$ we can now form
representations of the $SU(2)$ algebra
\eqn\rep{\Psi_{j,m}(z)\sim (J_0^- )^{j-m} \Psi_{j,j}(z),
~~~~m=-j,-j+1,...,j. }
The closed string vertex operators (with dimension $(1,1)$ and
ghost number two\foot{The theory also has non-trivial closed string
cohomology at ghost number $0$ that forms the ground ring,
and
at ghost number 1, which is associated with the conserved charges dual
to \wi $ $ \witten \ .}) in the theory are
\eqn\css{S_{j,m}(z, \bar z)=Y_{j,m}\bar Y
_{j,m},~~~~~~~Y_{j,m}=c\Psi_{j,m} e^{2i(1-j)\varphi}.}
The $S_{j,j}$'s  are the tachyon vertex operators that we have
already discussed. The rest  are new fields that are called the
discrete states. In our case this name is a bit misleading since
the tachyon modes are discrete as well.
This analysis differs from the usual $c=1$ case only by the
inclusion of appropriate factors of $i$. Again we emphasize that
this is a crucial  $i$  since it turns the discrete states into a
physical   propagating modes.

It is easy to see that the energy in the $X$ direction of
$S_{j,m}$ is $2m$ while the momentum in the $\varphi $ direction
is $2j$. Therefore, when computing \eg\ the $1\rightarrow
1$ amplitude we get, instead of \ec $ $ and \mc,
\eqn\fit{2 m^+  +2 m^- = 0,~~~~~~2-2g-(n+1+1)+{1\over 2}(2 j^+ + 2
j^-) =0, }
and so the sphere amplitude scales like
\eqn\spaa{ A(m^-, j^-; m^+, j^+)\sim \delta_{m^+ + m^-}~ \mu_0^{
j^+ + j^- }.}

On the quantum mechanics side we can  follow the same steps. The
$SU(2)$ generators are
\eqn\susu{J^{+} = \half \tr((a^{\dagger})^2), ~~~~J^{-}=
\half \tr(a^2),~~~~J^3={1 \over 4}\tr(a a^{\dagger} +a^{\dagger} a).}
The highest-weight states with respect to this $SU(2)$ algebra are
$\gamma_{j,j}=c_j~\tr ((a^{\dagger})^{2j}) $, where
$j=0,1/2,1,...$. Again we can form a representation of the algebra
by commuting $j-m$ times  $\gamma_{j,j}$ with $J^- ~ $ to get
$\gamma_{j,m}$. Up to ordering issues  that are not relevant for
the qualitative discussion of this section we find
\eqn\gag{\gamma_{j,m}\sim \tr( a^{j-m}(a^{\dagger})^{j+m} ). }
These operators satisfy a commutator algebra which at large $N$
approaches the semi-classical $w_\infty$ algebra \we. Even at
finite $N$, the algebra closes, using the gauge equivalence
\gauss. The precise details of the finite $N$ corrections to \we,
however, remain to be determined.

The two-point function associated with the discrete states takes
the form
\eqn\ovb{ {\cal O}(m^-,j^-;m^+,j^+)\sim \bra{0}\tr( a^{j^- +m^-}
(a^{\dagger})^{j^- -m^-})\tr(a^{j^+ -m^+} (a^{\dagger})^{j^+ +m^+}
)\ket{0}}
$$\sim \delta_{m^+ + m^-}~ N^{j^+ + j^-},$$
again in  agreement with the string theory result \spaa $ $.

\newsec{A closer look at the correlation functions}

In this section we consider some of the correlation functions
discussed in the previous section in more detail.
Let us again start with the harmonic oscillator. It is a simple
calculation to show
that
the $1\rightarrow 1$ amplitude \rtwop\ is
\eqn\rtwopp{ {\cal O}(k_1;k_2) = \delta_{k_1,k_2} k_1 N^{k_1}
\left( 1 + \CO(N^{-2}) \right).}
That follows from the fact that there are $k_1$ planar ways  to
commute the $a$'s through the $a^{\dagger}$.  Somewhat more
involved combinatorics show that for the $1\rightarrow 2$ overlap
amplitudes, \jk, the planar limit gives
\eqn\hjk{ {\cal O}(k;k_1,k_2) =\delta_{k, k_1+k_2} k k_1 k_2
N^{k-1}.}
The simplest way to calculate $1\rightarrow m$ overlap amplitudes
\eqn\onen{ {\cal O}(k;k_1,k_2,...k_m)=\bra{0} \tr(a^{k})\tr
((a^{\dagger})^{k_1})\tr ((a^{\dagger})^{k_2})...\tr
((a^{\dagger})^{k_m}) \ket{0},}
in the large-$N$ limit is to use the $w_{\infty}$ algebra $m$
times to move $\tr(a^{k})$ to the right. From \we \foot{Since we
are interested in the large-$N$ limit we can simply replace the
Poisson bracket in \we $ $ by a commutator.}$ $ we get
\eqn\onetn{ {\cal O}(k;k_1,k_2,...k_m)=\delta_{k, k_1+k_2+...k_m} k
(k-1)(k-2)...(k-m+2) k_1 k_2...k_m  N^{k-m+1} .}

To match this on the string theory side, we have to compute the
relevant sphere amplitudes in full detail. As explained in the
previous section the relevant calculation is equivalent to
scattering amplitudes in the free theory ($\mu_0 =0$) with extra
insertions of the screening operator. In particular, on the sphere
we get for the $1\rightarrow m$ tachyon scattering amplitude
\eqn\sct{ A(k;-k_1,-k_2,...,-k_m)_{\mu_0}={\mu_0^n \over n!}
A(k;0,0,...,0,-k_1,-k_2,...,-k_m)_{free}~, }
where the zeros indicate the insertion of the screening operator
$n=k-m+1$ times. In critical string theory one cannot write down a
closed formula for such  amplitudes. However, in 2D this can be
done using various approaches\foot{The motivation for studying
scattering amplitudes in this theory was as a tool for finding the
scattering amplitudes in the usual $c=1$ theory.} \refs{\PolyakovQX,
\KlebanovVP, \KlebanovUI, \DiFrancescoSS}. The result of the most general
$1\rightarrow r$ amplitude is
\eqn\onetor{ A(k;-k_1,-k_2,...,-k_r)_{free}=\prod_{i=1}^r
{\Gamma(1-k_i) \over\Gamma(k_i)} {\pi^{r-2} \over (r-2)!}. }
Combining this with \sct\ we find
\eqn\allequationshavethesamename{ 
A(k;-k_1,-k_2,...,-k_m)_{\mu_0}={\pi^{k-1}\over (k-m+1)!
(k-1)!} \left( {\mu_0 \Gamma(1) \over \Gamma(0)}\right) ^{k-m+1} ~
\prod_{i=1}^m {\Gamma(1-k_i) \over\Gamma(k_i)}.  }
This expression  contains some  zeros and some infinities that
must be dealt with before comparing with the harmonic oscillator
results. There are two kinds of zeros. The first comes from   the
$(1/\Gamma(0))^{k-m+1}$ and is due to the screening operators.
Recall that to obtain finite results
in the Liouville theory one needs to renormalize the cosmological
constant when $b\rightarrow 1$
(see \eg\ \TeschnerRV).  To be precise we need to take
 $\mu_0\rightarrow \infty$ and $b\rightarrow 1$ while keeping
\eqn\reg{\mu =\pi\mu_0 {\Gamma (b^2)\over \Gamma(1-b^2)},}
fixed. This is exactly the combination that appears in 
\allequationshavethesamename\ so
we can replace the $ \mu_0 { \Gamma(1) \over \Gamma(0)}$ in that
equation by $\mu/\pi$. A precise way to think about this procedure
is the following. Take $b=1+\epsilon/2 $ and twist the boundary
condition on the tachyon field $T(\varphi )=e^{2\pi i\epsilon
}T(\varphi +2\pi)$. That twisting has the effect of lifting the
tachyon  zero mode to $k=\epsilon$ so that the $\mu_0\Gamma(1)/
\Gamma(0)$ in \allequationshavethesamename\ becomes
$\mu_0\Gamma(1+\epsilon)/\Gamma(\epsilon)$ and in the limit
$\epsilon\rightarrow 0$ we simply get $\mu/\pi$.

The formula \allequationshavethesamename\ also vanishes
when one of the outgoing momenta, $-k_i$, is positive. This is
exactly as it should be since there are no operators on the
harmonic oscillator side that correspond to $T^{-}_{k>0}$ (see
discussion after eq.\ \ua).

Infinities arise from the $\Gamma$ functions
when $k_i>0$. The reason for these infinities is
that the amplitudes are sitting on a resonance. Physically there
is no region of interaction, the particles are always nearby and
they keep interacting forever. In the harmonic oscillator side
we avoided this by not propagating in time each of the operators.
To realize the same effect on the string theory side we need to
introduce leg factors. Heuristically, these can be viewed as
putting back the external legs that are missing in the
amputated S-matrix elements calculated by
string theory amplitudes.
The way to find the relevant leg-factors is to
match the two-point function on both sides of the duality. Then
one can test the duality by comparing the higher point functions.
The relevant leg-factors are\foot{The leg factors
for $k<0$ are familiar from the usual matrix model 2d string
duality (in Euclidean signature). We could have taken the same
leg-factors for $k>0$ (since $\Gamma(-n+\epsilon)={1\over \epsilon
\Gamma(n+1)}+{\cal O}(\epsilon^0)$) but that would have made the
relation with the matrix model a bit more clumsy. }
\eqn\leg{
f_{{\rm ket}}(k)=- {1 \over \pi} {\Gamma(k)\over \Gamma(-k)},
~~(k<0)~~~
 ~~{\rm and} ~~~
 f_{{\rm bra}}(k)=\pi \Gamma(k) \Gamma(k+1), ~~(k>0).
}
%
Putting all of this together one finds that  the relation between
the string theory scattering amplitudes and the matrix model
overlap amplitudes for the $1\rightarrow m$ processes is
%
\eqn\stmm{{\cal O}(k;k_1,k_2,...k_m)= f_{{\rm bra}}(k) 
\prod_{i=1}^m f_{{\rm ket}}(k_i)
A(k;-k_1,-k_2,...-k_m)_{\mu},}
with $\mu =N$. This relation holds in the $1\rightarrow 1$ cases
by construction. The fact that it holds for the other cases
($m>1$), that have non-trivial dependence on the $k$'s (see eq.\
\onetn) is  an encouraging consistency  check of the proposed
duality.

So far we did not consider the winding modes. Since $\varphi$ is
compactified, modular invariance implies that there must be winding
modes in the theory. But there are no candidate dual operators for
the winding modes in the quantum mechanics.\foot{The usual
suspects are the Wilson loops. But since $X$ is not compactified
there are no such operators in the gauged quantum mechanics.} This
is exactly the same puzzle we have encountered with the momentum
modes with opposite chirality. In that case the resolution was
that  scattering amplitudes with at least one opposite-momentum
mode vanish. The same is expected to happen for the winding modes.
As in the case of the momentum modes with opposite chirality,
we do not know how to prove this in general, but we demonstrate it\
here for some  cases. Consider, for example, scattering amplitudes of
$m$ winding modes. The analog of
\mc\ in that case is
\eqn\wind{ 2-2g-(n+m)=0,}
where again $n$ is the number of insertions of the screening
operator. We see that no amplitude with $m>2$ can satisfy
this  relation. For $m=1$ and  $m=2$ the relation can be satisfied
on the sphere with $n=1$ and $n=0$ respectively. But both cases
vanish since they  involve only two insertions of closed strings
on the sphere.

\subsec{Normal matrix models and matrix integral representation}

We end this section with a discussion on a  relation with normal
matrix models. A normal matrix model (NMM) is an integral over
complex matrices $Z$ such that $[Z, Z^\dagger] = 0$. The fact that
$Z$ commutes with its conjugate implies that, much like in the
case of a single matrix, in many interesting cases the model can
be described in terms of the eigenvalues of $Z$ and $Z^\dagger $.
The NMM that appears to be related to the quantum mechanics
described here takes the form
\eqn\normalmatrixint{ \CZ_{\rm NMM}(J, J^\dagger) = \int_{[Z,
Z^\dagger] = 0 }  d^{N^2}Z d^{N^2}Z^\dagger ~  e^{  - \tr
(Z^\dagger Z+ Z J^\dagger + Z^\dagger J ) }. }
%
The normal-matrix
constraint in the path integral
can be imposed  with a
Lagrange multiplier matrix $A$,
\eqn\deltarep{ \delta^{N^2}\left( i [ Z,
Z^\dagger] \right) = \int dA ~e^{ \tr [Z, Z^\dagger] A }, }
so that \normalmatrixint\ becomes
\eqn\normalmatrixintagain{ \CZ_{\rm NMM}(J, J^\dagger) = \int d^{N^2} A
d^{N^2}Z d^{N^2}Z^\dagger ~  e^{  - \tr (Z^\dagger Z+ [Z,
Z^\dagger] A+ Z J^\dagger + Z^\dagger J ) }. }
We recognize the first two terms in the exponent as the
Hamiltonian of the gauged harmonic oscillator (when one does not
fix the gauge $A=0$). This strongly indicates a relation with the
quantum mechanics, where the NMM should be viewed as the
Hamiltonian reduction of the quantum mechanics. One expects to be
able to compute (at least in the large-$N$ limit) the overlap
amplitudes discussed above using the NMM. In particular, by
comparing Wick contractions, one should find a relation of the
form \eqn\matrixsource{ \prod_i \tr \left(\del_J^{k_i}
\del_{J^\dagger} ^{p_i} \right) |_{J=J^\dagger =0} \ln \CZ_{\rm
NMM}(J, J^\dagger) = \vev{ T \left( \prod_i \tr a^{\dagger~k_i}
a^{ p_i } \right)}} where $T$ indicates a certain ordering
prescription.

Here we mention one interesting way  to think about this connection.
Recall the Wigner phase-space integral representation of
expectation values in one-dimensional quantum mechanics:
\eqn\wignerintegral{
\bra{\psi} \hat{\CO} \ket{\psi} = \int dx dp~W_\CO (x,p)
W^\star_\psi(x,p), }
where \eqn\weylrep{ W_\CO(x,p) \equiv \int dy
~e^{ i p y} \bra{ x + { y\over 2}} \hat \CO \ket{ x - {y \over 2}
 } ,}
 and the Wigner function of the state $\psi$ is
\eqn\wignerfunction{ W_\psi(x,p) \equiv W_{ \ket{\psi}\bra{\psi}}
=\int dy ~ e^{ i p y } \psi^\star \left(x+ {y \over 2}\right)
\psi\left(x-{y \over 2}\right).}
For the case of the harmonic oscillator this representation is
particularly interesting. To compute the  vacuum expectation
values of some operator we simply use the fact that the ground
state wave function is
$ \psi(x) = \psi_0(x) = e^{ - x^2/2} $, and eq.\ \wignerintegral\
becomes \eqn\harmonicwignerintegral{ \bra{\psi_0} \hat \CO
\ket{\psi_0}= \int dx dp~e^{-H(x,p)}~W_\CO(x,p). } This quantum
expectation value is equal to a classical statistical average at
inverse temperature $\beta = 1$, the resonant frequency. The
generalization of these formulae to $U(N)$ matrix quantum
mechanics look very similar to \normalmatrixint. There are,
however, various subtleties involving ordering (related
to the ordering in \matrixsource) and gauge fixing that should
be clarified.

\newsec{Beyond tree level}

So far on the string theory side we have considered in detail only
sphere amplitudes. In this section we first test our proposal at
the one-loop level and then discuss some non-perturbative aspects.

\subsec{Torus vacuum energy}

Let us consider the simplest one-loop amplitude. This is the
one-loop contribution to the ground state energy.  On the string
theory side the ground state energy is the expectation value of
the zero momentum graviton which is one of the discrete states
discussed in section 5, $S_{1,0}$. The energy conservation
condition is automatic (since $m=0$) and the $\varphi$ momentum
conservation gives
\eqn\qas{n=2-2g. }
This means that there are two possible perturbative contributions.
The first
comes from the sphere and since $n=2$ it scales like $N^2$. The
second comes from the torus with $n=0$, and it
therefore scales like $N^0$.

On the quantum mechanics side the ground state energy is $N^2/2$
with no additional constant that scales like $N^0$. This suggests
that  the one-loop vacuum energy in this string theory  should be
zero. This seems unlikely since this is a bosonic string theory
and it is hard to see what could possibly cancel the positive
vacuum energy. On the other hand this is a somewhat unconventional
string theory and so it is worthwhile to do the calculation and
see what we get.

A simpler way to compute the ground state energy is to calculate
the one particle irreducible contribution to the partition
function with $X$  compactified, $ X\sim X + \beta.$ In the limit
$\beta\rightarrow \infty$ we have
\eqn\er{\ln Z= -\beta E_0,}
from which we can read off $E_0$. Thus what we have to do is to
compute the torus partition function.
%
%
That calculation is relatively simple since the background charge
on the torus is  zero and so it does not involve
the Liouville
term at all.
As far as this calculation is concerned $X$ and
$\varphi$ are two free scalar fields. In fact this calculation was
already done in the context of the usual $c=1$ theory
\refs{\GuptaFU, \SeibergEB, \BershadskyZS}.  The
motivation for that calculation was to study the theory at finite
temperature. In that case the Liouville direction is not
compactified and $X$ is compactified $X\sim X+2\pi R$. The result
of that calculation is \BershadskyZS
\eqn\pqw{ {Z_{torus}\over V_L}={1\over 12\sqrt{2}} \left(
{R\over\sqrt{\ap}}+{\sqrt{\ap}\over R}\right) ,}
where  $V_L$ is the volume in the Liouville direction.

Since in that calculation $X$ and $\varphi$ are two free scalars
we can interchange their role to find the partition function in
our case. Interchanging their role means that  now $\varphi $ is
compactified and
\eqn\oa{ V_L \rightarrow V_X =\beta,}
which gives
\eqn\tpp{ \lim_{\beta\rightarrow\infty} {Z_{torus}\over \beta}=-{
i\over 12\sqrt{2}}\left( {R\over\sqrt{\ap}}-{\sqrt{\ap}\over
R}\right) .}
The $i$ and the minus sign come from the fact that $\varphi$ is a
timelike direction. Another way to see this is to take $\ap
\rightarrow -\ap$ in \pqw $ $. Since in our case $R=1$ in units
where $\ap=1$\foot{Since $\varphi$ is time-like this is not the
self-dual radius. Acting on a time-like direction, T-duality takes
$R$ to $-\ap / R$, in order that the partition function be
invariant.} we find that
\eqn\tq{ \lim_{\beta\rightarrow\infty}{Z_{torus}\over \beta}=0,}
as predicted by the duality with the quantum mechanics.

This is an interesting result for several reasons. First, it is
obtained by summing over contributions of all of the physical
states of the string. In addition to the chiral momentum modes
with nonzero S-matrix amplitudes, these include winding modes and
momentum modes with the opposite chirality that, we argued,  do
not appear on external legs. This is reminiscent of the ghost
structure of loop amplitudes in a gauge theory, and one is tempted
to attribute this behavior to the underlying $w_\infty$ symmetry.
Second, one should always take notice when finding zero for the
quantum correction to the cosmological constant, especially in a
bosonic theory.

\subsec{Hints for nonperturbative effects}

The fact that the perturbative expansion for all scattering
amplitudes is truncated seems to imply that there is no room  for
non-perturbative effects in this theory. Indeed instantons are
usually related to the non-convergence of the perturbative
expansion, which is not an issue with finitely many terms. On the
string theory side this means that there are no D-instantons.
Below we provide some evidence, from the quantum mechanics side,
that there are, however, D-particles in the theory.

Consider the matrix model at finite temperature. Assume that the
temperature is low so that the contribution of the zero modes of
the gauge fields can be neglected.  The  free energy is then
\eqn\partition{ F(q,N) = \ln Z_N(q) = \half \beta N^2 -
\sum_{n=1}^N \ln ( 1 - q^n),~~~~~~q=e^{-\beta} . }
 Keeping track of the
$N$-dependence, this can be rewritten (a similar equation appears
in $QCD_2$ \GrossTU) as
\eqn\partitionN{ F\left(q,N \right) = {\beta N^2\over 2 } +
f(\beta)+\sum_{n=1}^{\infty} C_n \left(e^{ -n \beta N}
\right),~~~~~~f(\beta)=\sum_{n=1}^{\infty}\ln( 1 - q^n) . }
The first term we recognize as the ground state energy. The second
term is associated with the leading  torus contribution to the
free energy when $\beta $ is large but finite. There are no
further perturbative corrections in accord with the
scaling rule \emc.
There are, however, non-perturbative effects. The fact that these
scale like $ e^{-w\beta N}$ implies that there are D-particles in
the spectrum which contribute to the partition sum when their
worldlines wrap the thermal circle. The fact that there are no
terms of the form $e^{- N} $ implies that, as was argued above,
there are no D-instantons. It would be very interesting to verify
this statement using worldsheet techniques.

Another reason to study D-branes in this theory is the following.
We motivated our search for the string dual of the harmonic
oscillator by noticing that reversing the sign of the string
tension in the usual $c=1$ theory
rights the upside-down oscillator.
This identification of $-{1\over \alpha'}$ with
the mass of the open-string tachyon
relies on the D-brane physics of the
2d string theory.
The analogous understanding for this
'imaginary Liouville' theory remains to be
developed.
The branes which
describe the matrix eigenvalues
in this case
are closely related to the
instantonic branes of the
usual $c=1$ theory \KlebanovKM,
which experience trajectories of the form
$ z = \tilde \lambda \cos t $.
We note that the gauging of the $U(N)$ symmetry
should again be motivated by the presence of a
null open-string descendant.
A more detailed understanding
of these issues will allow a
clear intepretation of the relation discussed
in this paper as
open/closed duality
along the lines of \refs{\McGreevyKB, \KlebanovKM, \McGreevyEP}.

\newsec{The worldsheet as a phase-space}

So far we have presented quite a bit of evidence that supports the
proposed duality. What is still missing is a simple
intuitive picture of how the duality works. Below we give a
heuristic description of how, we believe, the closed string
world-sheet is related to the harmonic oscillator phase-space.

It is reasonable that $\mu$ on the string theory side is actually
the conjugate variable to $N$. Namely, it is the chemical
potential. At large $\mu$, these agree $\vev{\hat N} = \mu$. There
are two reasons to suspect that this is the case. First, that is
the way the usual duality between 2d strings and the matrix model
works. Second, at the moment, it is not clear what effect on the
string theory side could force $\mu$ to be an integer.\foot{This
argument is not too convincing since there is a  counterexample
to the reasoning, namely the duality of \GopakumarKI.  In this
topological case, the open-string side of the duality implies that
the closed string coupling constant should be quantized. But the
closed string dual seems to make sense for all values of $g_s$.}
Assuming that this is the case we see that something quite
interesting is happening. From the world-sheet point of view $\mu$
is the conjugate variable to the area of the world-sheet. From the
target space point of view it is conjugate to $N$ which is the
area of the phase-space. This seems to suggests that the
worldsheet should be related (via some non-local map) with the
phase-space of the harmonic oscillator.

At first sight this seems unlikely since the phase-space has a
boundary while the closed string word-sheet does not. However,
recall  the nature of the calculations we did on the quantum
mechanics side that had a simple interpretation on the string
theory side. We were computing overlap amplitudes
\eqn\olj{\vev{{\rm bra| ket}}.}
At the semi-classical level this is equivalent to  exciting  two
copies of the Fermi sea, and gluing them together along their
Fermi surfaces to form a closed manifold, which could be viewed as
the closed-string worldsheet (see fig. 2).
\ifig\gluing{ Gluing together two copies of the phase space along
the Fermi surface produces a closed Fermi sea, which could be
viewed as the closed string world sheet. }
{\epsfxsize1.5in\epsfbox{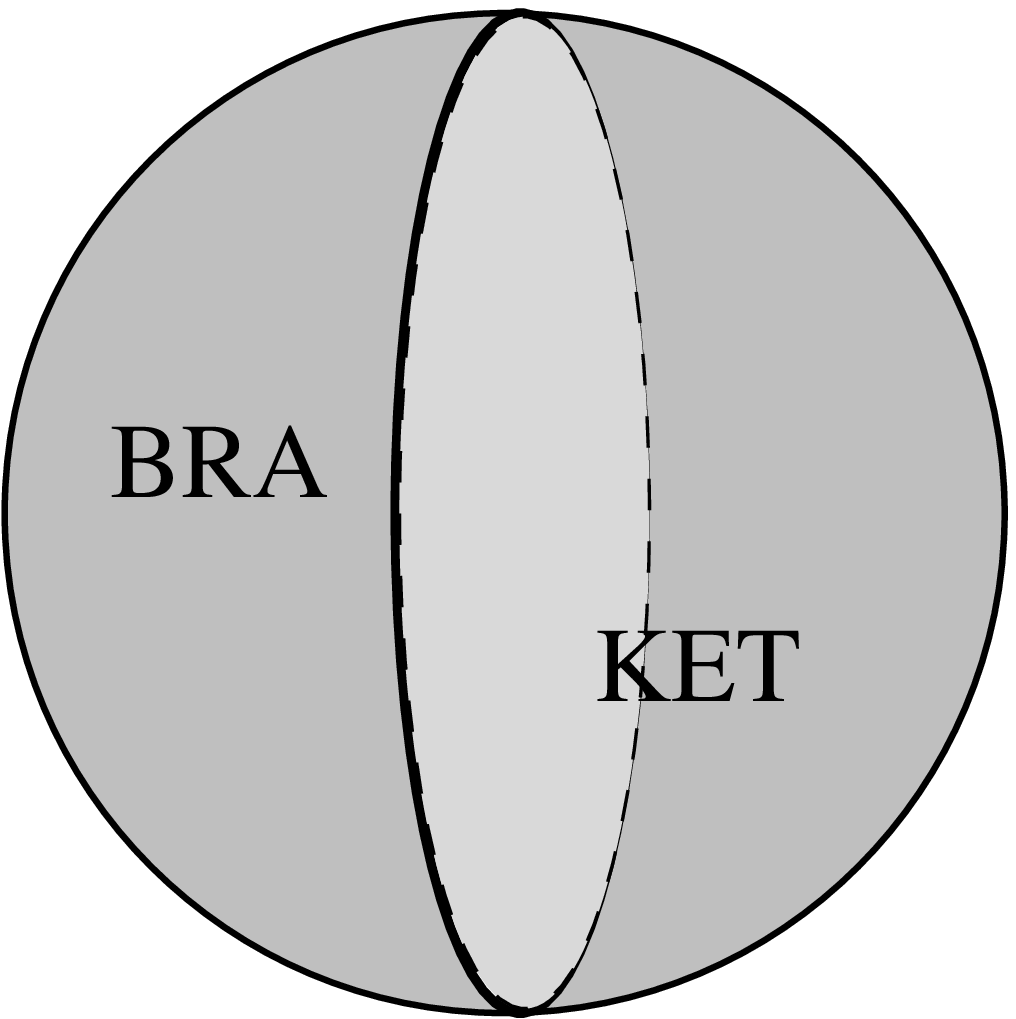}} This closed two-dimensional
space might be interpreted as the closed string world-sheet. The
analog of the target-space energy conservation condition on the
world-sheet is the level matching between the left and right
movers
\eqn\utr{E_{\rm bra}=E_{\rm ket} ~~ \Leftrightarrow ~~ N_L = N_R
.}
It should be interesting to see how precise this can be made. In
particular, it would be nice to make  contact with \rajesh $ $
where it was illustrated how  a  closed string world-sheet could
be realized in the large-$N$ limit of a free field theory.

\newsec{Discussion }

There are various
generalizations and applications of the duality
proposed here that might be interesting to study. Here we
mention some of them.

\subsubsec{$c<1$}

The  simplest generalization of the string theory described here
(other than the supersymmetric generalization) is to consider
minimal models (with $c<1$) times a time-like Liouville direction
with an imaginary linear dilaton so that the total central charge
is $26$. One way to think about this theory is to start with the
usual minimal strings where the Liouville direction is space-like
and the linear dilaton is real and Wick rotate the Liouville
direction.\foot{ This is different than taking $\ap\rightarrow
-\ap$ since we do not "Wick rotate" the minimal model.} That
theory is quite amusing  since the minimal models have a Coulomb
gas description in terms of
a space-like scalar with an imaginary linear dilaton. So we end up
with  a string theory in 2D Minkowski space-time with an imaginary
dilaton that depends on a certain linear combination of the time
and the space directions. The string theory considered here is a
special case in which the imaginary linear dilaton depends only on
the time-like direction, $\varphi$. The fact that the $\varphi $
direction is compact prevents one from relating the theories by a
boost.  It is likely that these string theories are dual to some
matrix models. The results of the present paper seem to indicate
that these matrix models should not involve double scaling. It
should be interesting to explore the relation between these
theories and the usual minimal strings, especially in light of
recent progress (see \eg \ \refs{\SeibergNM,\GaiottoYB}).

\subsubsec{Topological strings}

The fact that the perturbative corrections to the amplitudes in
this string theory truncate after a finite number of terms may
seem to conflict with one's intuition about unitarity. This
confusion is resolved however, when one realizes that these
amplitudes do not have an imaginary part. In fact, as is obvious
on the matrix model side, they are integers that are equal to the
number of diagrams with given topology, fixed by the
external lines and the genus. This makes a connection with
topological string seem inevitable. Given that deep connections
have been found between noncritical strings and topological
strings on noncompact Calabi-Yaus (\eg\ \refs{\AganagicQJ,
\GhoshalWM}) and further that 2d Yang-Mills also has a topological
string description \VafaQA, it seems likely that there is a
topological string description of the matrix harmonic oscillator.

\subsubsec{Stringy compactification }

The fact that the string theory considered in this paper seems to
makes sense despite its strange properties suggests that
generalizations to higher dimension might be interesting to
explore. Perhaps these could even lead to a new class of stringy
compactification with phenomenological applications. For example
consider string theory on $\IR ^d \times S^1$ where the radius of
the $S^1$ is $R$. The usual phenomenological problem with such a
compactification is that the radius of the $S^1$ is not fixed and
as a result the $d$ dimensional physics contains a massless scalar
that (under reasonable assumptions) contradicts observation.
Fixing such moduli is a non-trivial problem in string theory. Here
we seem to encounter  a simple way to deal with this: Suppose that
we turn on an imaginary linear dilaton in the $S^1$ direction. As
discussed in section 4, periodicity of the string coupling
constant fixes $R$ to be an integer $m$. For $m\gg 1$ we find that
the contribution of that direction to the central charge is almost
$1$. And so the difference, which scales like $1/m^2$, could
easily be cancelled against some other fluxes in the theory
without involving significant  $\ap $ corrections.

Needless to say, one should be careful here. The fact that in 2d
having a complex string coupling constant seems to work does not
mean that the same will happen in higher dimensions. In
particular, in 2d only the zero mode of the dilaton appears in the
cohomology, $S_{1,0}$, which makes the unitarity condition much
less restrictive.

\subsubsec{The quantum Hall effect}

The matrix model studied in this paper is closely related to the
quantum hall effect (for a beautiful review of the QHE see
\girvin). The simplest way to see this is at the level of the
classical phase-space. If we replace $ P $ by $ B_z Y$ (where
$B_z$ is the magnetic field and $Y$ is interpreted as a new
spatial direction), we end up with the familiar cyclotron motion
around the origin $X=Y=0$. The fact that the center of the
cyclotron motion is fixed means that translation invariance in the
$X, Y$ plane is broken, which is crucial for realizing the QHE .
Namely, this is the analog of the Anderson localization in the
presence of an impurity located at the origin of a sample in the
shape of a  disk. At the quantum level, this is clearest when
considering the problem in the symmetric gauge $\vec{A}=-{1\over
2}\vec{r}\times\vec{B}$.  In that gauge it is convenient to label
states by their energy $n+1/2$ (at the nth Landau level)
and their angular momentum $m$.\foot{Recall that due to the
large magnetic field  $m$ cannot be negative.
This translates to the chiral nature of the tachyon in our case.
} There is no further
degeneracy in the problem. The states we have in the harmonic
oscillator correspond to $n=m$, which means that in each Landau
level we have exactly one state.

A long-standing problem in the QHE is to find a CFT description of
the quantum phase transition from one plateau to another. The
jump from one plateau to the next is associated with occupying a
new Landau level. The analogous transition in our case is of
adding a new eigenvalue to the system, which means on the string
theory side adding a D-brane. This might be viewed as an
indication that the  CFT that describes  the phase transition in
the QHE is the one considered here, but on the disk rather than on
the sphere. This fits neatly with the heuristic picture of the
previous section. Now the worldsheet has a boundary that is
associated with the disk shaped sample. All the interesting
physics associated with the edge states should correspond to open
strings inserted at the boundary of the worldsheet. The fact
that much progress has been made recently in understanding the open
string spectrum of D-branes in the usual Liouville theory
\refs{\FateevIK, \TeschnerMD, \ZamolodchikovAH}
suggests
that this
speculation could be tested in the near future.

\bigskip
\centerline{\bf{Acknowledgements}}

We thank A. Hashimoto, V. Oganesyan, L. Rastelli, E. Verlinde and
especially I. Klebanov for discussions.
JM is supported by a Princeton University Dicke
Fellowship, and by the Department of Energy under Grant No.\
DE-FG03-92ER40701,
and would like to thank
the Aspen Center for Physics for hospitality during the course of
this work.
NI is  supported in part by the National
Science Foundation under Grant No.\ PHY 9802484. Any opinions,
findings, and conclusions or recommendations expressed in this
material are those of the author and do not necessarily reflect
the views of the National Science Foundation.

 \listrefs
\end

The amplitudes we are studying can be computed by a
zero-dimensional matrix model. This is an auxiliary matrix
integral with matrix sources: \eqn\matrixintegral{ \CZ(J,
J^\dagger) = \int d^{N^2}Z d^{N^2}Z^\dagger ~
e^{ -\tr ( Z^\dagger Z  + Z J^\dagger + Z^\dagger J ) } = e^{ \tr
J J^\dagger}. } Correlators of the matrix quantum mechanics
satisfy
\eqn\matrixsource{ \prod_i \tr \left(\del_J^{k_i}
\del_{J^\dagger} ^{p_i} \right) |_{J=J^\dagger =0} \ln \CZ(J,
J^\dagger) = \vev{ T \left( \prod_i \tr a^{\dagger~k_i} a^{ p_i }
\right)}} where $T$ indicates a certain ordering prescription
which will become clear momentarily.

There are two ways to arrive at the result \matrixsource. A crude
explanation for \matrixsource\ is that the propagator needed to
compute the correlators in the matrix quantum mechanics
is just the same as the propagator
in the zero-dimensional matrix integral. So the two calculations
involve the same combinatorics of Wick contractions, with the same
2-point function\foot{ It is very tempting to suggest that this
description is related to the normal matrix model of
\AlexandrovQK.  This is similar to \matrixintegral, but with the
measure of $Z$ restricted to normal matrices, $[Z, Z^\dagger] =
0$, \eqn\normalmatrixint{ \CZ_{\rm NMM}(J, J^\dagger) = \int d^{N^2}Z
d^{N^2}Z^\dagger ~ \delta^{N^2}([Z, Z^\dagger])~ e^{ -\tr (
Z^\dagger Z  + Z J^\dagger + Z^\dagger J ) }. } Imposing the
normal-matrix contraint with a Lagrange multiplier matrix $A$,
\eqn\deltarep{ \delta^{N^2}\left( i [ Z, Z^\dagger] \right) = \int
dA ~e^{ \tr [Z, Z^\dagger] A }, } it is worth noting that
$H(Z,Z^\dagger,A)$ appearing in
$$ \CZ_{\rm NMM} = \int dA dZ dZ^\dagger
e^{-H(Z, Z^\dagger, A)} $$ is the Hamiltonian of the MQM. Defining
the inverse propagator \eqn\propG{ G(A)^{-1} = \Ione + \ad_{A}, }
and integrating out $Z$, we find \eqn\generating{ \CZ(J,
J^\dagger) = \int dA~\det{}^{-1}(G(A)) ~e^{\tr J^\dagger ( G(A) J)
} } Therefore \eqn\contract{ \prod_i \tr \del_J^{n_i}
\del_{J^\dagger}^{m_i} \CZ = \int dA ~\det{}^{-1}(G(A)) \prod_i
\tr \del_J^{n_i} (G(A)J)^{m_i} .} The contributions of $A$ are
always of the form $\tr \ad \cdots = 0$, and therefore this
reduces to the contribution from $ G(A = 0)$, for which we may use
the generating function \matrixintegral, \ie\ the propagator is
identical to the MQM propagator. It would be interesting to
understand the connection with \AlexandrovQK\ in more detail. }

The only subtlety is that the matrix integral computes the
correlator with a particular ordering. This ordering can be
understood as follows. It is possible to show that the LHS of
\matrixsource\ is the Wigner phase-space integral representation
of the correlator on the RHS.

Recall first the phase-space integral representation of
expectation values in one-dimensional quantum mechanics:
\eqn\wignerintegral{ \bra{\psi} \hat{\CO} \ket{\psi} = \int dx
\int dp~W_\CO (x,p) W^\star_\psi(x,p) } where \eqn\weylrep{
W_\CO(x,p) \equiv \int dy ~e^{ i p y} \bra{ x + {\hbar \over 2}y}
\hat \CO \ket{ x - {\hbar \over 2} y } } and the Wigner function
of the state $\psi$ is \eqn\wignerfunction{ W_\psi(x,p) \equiv W_{
\ket{\psi}\bra{\psi}} =\int dy ~ e^{ i p y } \psi^\star \left(x+
{\hbar \over 2}y\right) \psi\left(x-{\hbar \over 2}y\right).} For
the case of the harmonic oscillator this representation is
particularly interesting. For vacuum expectation values,
$ \psi(x) = \psi_0(x) = e^{ - x^2/2} $, \wignerintegral\ can be
rewritten as \eqn\harmonicwignerintegral{ \bra{\psi_0} \hat \CO
\ket{\psi_0}= \int dx dp~e^{-H(x,p)}~W_\CO(x,p) } This quantum
expectation value is equal to a classical statistical average at
inverse temperature $\beta = 1$, the resonant frequency.

For the non-gauged matrix oscillator, these formulae can be
directly matricized. This implies that the phase-space-integral
representation of the non-gauged matrix harmonic oscillator
correlators are generated by the integral \matrixintegral.
If the operator $\CO$ is Weyl-ordered in its matrix indices, the
matrix function $W_\CO$ will have the same form.

The gauging has the following effect.
The identity operator on the physical, gauge-invariant Hilbert
space is \eqn\identityop{ \Ione = \int d^{N^2} X \ket{X}\bra{X}
\hat\CP } where $ \hat \CP = \hat \CP^2 $ is the projection
operator onto the gauge-invariant subspace, \eqn\projectorop{ \hat
\CP = \int dA ~e^{ i \tr A [ \hat X, \hat P] } . } Acting on an
$X$ eigenstate, this integrates over its gauge orbit:
\eqn\translate{ \hat \CP \ket{X} = \int dA~\ket{ e^{iA} X e^{
-iA}}.} and the identity operator may be rewritten
\eqn\identityagain{ \Ione = \int dX \int dA \ket{X} \bra{ e^{ -i
A} X e^{ i A} }. }
The matrix integral expectation value for a Weyl-ordered operator
$\CO$ is then
\def\vvev#1{\langle\langle#1\rangle\rangle}
\eqn\mmm{ \vvev{\CO}= \int d^{N^2}X~d^{N^2}P
~W_{\psi_0}(X,P)~ W_\CO(X, P) .} This is \eqn\mmagain{ \int
dX dY_1 dY_2 dP ~e^{ i \tr P ( Y_1 - Y_2)}
\vev{\psi_0 | X + {\hbar \over 2} Y_1  } \bra{ X + {\hbar \over 2}
Y_2 } \hat \CO \ket{ X - {\hbar \over 2} Y_2 } \vev{ X - {\hbar
\over 2} Y_1 | \psi_0} } If the operator $\CO$ is gauge-invariant,
\eqn\gaugeinvariance{ \hat\CO(X)= \hat \CO(e^{iA}Xe^{-iA}) = \hat
\CP \hat \CO \hat \CP,} \eqn\awef{ \vvev{\CO} = \bra{\psi_0} \int
dX_+ \ket{ X_+  } \bra{ X_+ } \hat \CP \hat \CO \int dX_- \hat \CP
\ket{ X_- } \vev{ X _- | \psi_0} } we see in \awef\ two
resolutions of unity which we can remove to obtain $
\vvev{\CO}=\vev{\CO}$. The end result of this discussion is that
expectation values of gauge-invariant and Weyl-ordered MQM
operators may be computed using the formula \matrixsource. Note
that because of the gauging, one can represent every MQM operator
in terms of the ordering $\tr \del_J^m \del_{J^\dagger}^n $.

The phases arising from the complex dilaton conspire
to give this selection rule.

We have learned the following
lesson
about the liouville dressings here:
in this 'imaginary' version of Liouville,
the Seiberg bound becomes the fact that
the target space field is chiral.

The discrete states, which previously
had unphysical imaginary momenta
now become real modes of the theory.

Later, using the corley et al tricks \CorleyZK

If we include multiple flavors of oscillator,
the matrix model is just as solvable.
There is a hagedorn density of single-string states.
What is the string theory then?

\subsec{normal matrix model}

\newsec{Introduction}

A great deal has been learned from matrix models of strings in few
dimensions. The reinterpretation as open-closed string duality
puts these relationships in a modern perspective, from which we
might hope to [\eg\ find the right framework to prove
higher-dimensional open-closed dualities]. In an important
respect, however, the known exact descriptions of low-dimensional
strings by double-scaled matrix models are not precise analogs of
the description of strings in $AdS_5\times S^5$ by the $\CN=4$
SYM. The number of branes {\it must} be infinite in these models
to have a continuum string description. That is, the feynman
diagrams of double-scaled models discretize the worldsheet of the
string, rather than triangulating the moduli space of this
worldsheet. [Notable exceptions are the Kontsevich model
\KontsevichTI -- as explained in \GaiottoYB, the diagrams of this
model are the cubic OSFT diagrams -- and [maybe] the normal matrix
model \AlexandrovQK, about which see below.] In contrast, AdS/CFT
can be formulated at finite $N$.

The model we will discuss in this paper is the quantum mechanics
of a matrix harmonic oscillator,
$$ S = \half \int dt ~ \tr \left( (D_0 X)^2 - X^2 \right) .$$
The derivative $D_0 = \del_0 + [A_0, . ]$ is covariant with
respect to gauged $U(N)$ conjugations $ X \to \Omega X
\Omega^\dagger $. Very closely related models have been considered
\refs{\HashimotoZP, \CorleyZK, \Berenstein} as toy models for the
$\CN=4$ SYM theory. This model has two imporant features for our
purposes.  Firstly, it is ridiculously simple and completely
solvable.  Secondly, there is no critical behavior.  The only
dimensionless coupling is $N$. In the spirit of \rajesh\ we will
attempt to find a string theory description of this free gauge
theory at finite, large $N \sim {1\over g_s}$, with no
double-scaling involved. This is like AdS/CFT, except that there
is no analog of the 't Hooft coupling. This will allow us to probe
finite-$N$ effects, such as stringy exclusion.
Further, we will
see that at finite string coupling, there is a UV cutoff on the
target space momentum. A similar phenomenon of a target space
lattice with spacing of order $g_s$ has been observed in the
context of topological strings \OkounkovSP.

The string dual of such a simple field theory might be expected to
have some kind of dangerous pathology. And indeed, our candidate
dual is indeed apparently insane (the action isn't real, the
string tension is negative). It is, however, calculable, and to
the extent that it is the same as the harmonic oscillator, it is
well-behaved.

Another important fact to observe is that this string theory can
be studied experimentally.